# Bonding mechanism in the nitrides Ti$_2$AlN and TiN: an experimental and theoretical investigation

M. Magnuson[1,2], M. Mattesini[3], S. Li[1], C. Höglund[2], M. Beckers[2], L. Hultman[2] and O. Eriksson[1]

[1]*Department of Physics, Uppsala University, P. O. Box 530, S-751 21 Uppsala, Sweden.*

[2]*Department of Physics, IFM, Thin Film Physics Division, Linköping University, SE-58183 Linköping, Sweden.*

[3]*Departamento de Física de la Tierra, Astronomía y Astrofísica I, Universidad Complutense de Madrid, E-28040, Spain.*

**Abstract**

The electronic structure of nanolaminate Ti$_2$AlN and TiN thin films has been investigated by bulk-sensitive soft x-ray emission spectroscopy. The measured Ti *L*, N *K*, Al *L$_1$* and Al *L$_{2,3}$* emission spectra are compared with calculated spectra using *ab initio* density-functional theory including dipole transition matrix elements. Three different types of bond regions are identified; a relatively weak Ti 3*d* - Al 3*p* bonding between -1 and -2 eV below the Fermi level, and Ti 3*d* - N 2*p* and Ti 3*d* - N 2*s* bonding which are deeper in energy observed at -4.8 eV and -15 eV below the Fermi level, respectively. A strongly modified spectral shape of 3*s* states of Al *L$_{2,3}$* emission from Ti$_2$AlN in comparison to pure Al metal is found, which reflects the Ti 3*d* - Al 3*p* hybridization observed in the Al *L$_1$* emission. The differences between the electronic and crystal structures of Ti$_2$AlN and TiN are discussed in relation to the intercalated Al layers of the former compound and the change of the materials properties in comparison to the isostructural carbides.

## 1 Introduction

Ternary carbides and nitrides M$_{n+1}$AX$_n$ (MAX phases), where n=1, 2 and 3 refers to 211, 312 and 413 crystal structures, respectively, have recently been the subject of much research [1, 2, 3]. M denotes an early transition metal, A is a p-element, usually belonging to the groups IIIA and IVA, and X is either carbon and nitrogen [4]. These nanolaminated materials exhibit a technologically important combination of metallic and ceramic properties, with high strength and stiffness at high temperatures, resistance to oxidation and thermal shock, in addition to high electrical and thermal conductivity [5]. The macroscopic properties are closely related to the underlying electronic and crystal structures of the constituent elements and their stacking sequence. The family of MAX-phase compounds, with more than 50 energetically stable variants, has a hexagonal crystal structure with near close-packed layers of the M-elements interleaved with square-planar slabs of pure A-elements, where the X-atoms (C or N) fill the octahedral sites between the M-atoms. The A-elements are located at the center of trigonal prisms that are larger than the octahedral X sites.

The 211-crystal structure was derived in the early 1930:th when these materials were referred to as Hägg phases with certain stability criteria depending on the ratio of the radii of the constituent atoms [6]. The recent improvements in synthetization processes has led to a renaissance of these compounds with the discovery of the unique mechanical and physical properties and the refined single crystal thin film processing techniques [5, 7].

The Ti-Al-N ternary systems include $Ti_2AlN$ (211) and $Ti_4AlN_3$ (413). These materials have been known in their bulk form since the 1960s. Recently, single crystal thin films were synthesized [8], which provide better opportunities to determine their electronic structure properties. Intercalation of Al monolayers into the TiN matrix implies that the strong Ti-N bonds are broken up and replaced by weaker Ti-Al bonds with a cost of energy. Thus, in $Ti_2AlN$, every second single monolayer of N atoms in TiN have been replaced by an Al layer, in effect resulting in understoichiometric TiN. The $Ti_2N$ slabs surrounding the Al monolayers are then twinned with the Al layers as mirror planes. Figure 1 shows the crystal structure of $Ti_2AlN$ with thermodynamically stable nanolaminates of binary Ti-N-Ti layers separated by softer Ti-Al-Ti layers with weaker bonds [9]. As shown in Fig. 1, the 211 crystal structure contains $Ti_{II}$ atoms with chemical bonds both to the N and the A-atoms while stoichiometric TiN contains $Ti_I$ atoms which only bond to N. The chemical bonding contains a mixed contribution of covalent, metallic and ionic character where the strength of the covalent contribution is slightly different for TiN and TiC.

The material's elastic properties depend on X (C or N) and crystal structure. The Young's modulus ($E$) of single-crystal films of $Ti_2AlN$ (270 GPa [8]) is higher than for $Ti_2AlC$ (260 GPa [10]) which are both significantly lower than for the corresponding binary compounds TiN (449 GPa [11]) and $TiC_{0.8}$ (388 GPa) [12]. On the contrary, the hardness of $Ti_2AlN$ (16 GPa [8]) is lower than for $Ti_2AlC$ (20 GPa [10]) and comparable to the case of TiN (21 GPa [11]) and $TiC_{0.8}$ (30 GPa [12]). The change of elastic properties with X is mainly related to the additional valence electron in N and the larger electronegativity compared to C. The weak Ti-Al bonds also affect the tribological properties, such as wear performance and friction [5]. The physical properties of crystallographically oriented thin films of MAX phases provide opportunities for particular industrial applications such as wear protective coatings on cutting tools and diffusion barriers in contact materials in micro- and nanoelectronics.

Previous experimental investigations of the occupied and unoccupied electronic structure of $Ti_2AlN$ and TiN include valence band photoemission (VBPE) [14] and soft x-ray absorption (SXA), spectroscopy [15]. However, these methods are rather sensitive to surface contamination. In addition, SXA is hampered by significant core hole effects in the final state for both C and N. Due to the lack of dipole selection rules in Auger spectroscopy, the N $KL_{2,3}L_{2,3}$ and Ti $L_3M_{2,3}M_{2,3}$ lines directly overlap in TiN [16, 17]. Theoretically, it has been shown by bandstructure calculations that there should be significant differences between the valence-band partial density-of-states (pDOS) of Ti, N, C and Al of $Ti_2AlN$, $Ti_2AlC$, TiN and TiC [18, 19, 20, 15]. In recent studies of carbides, applying soft x-ray emission (SXE) spectroscopy, we investigated the three 312 phases $Ti_3AlC_2$, $Ti_3SiC_2$ and $Ti_3GeC_2$ [21], the 413 phase $Ti_4SiC_3$ [22] and the 211 phase $Ti_2AlC$ compared to TiC [23]. In contrast to $Ti_3SiC_2$, $Ti_3GeC_2$ and $Ti_4SiC_3$, a pronounced peak at -1 eV below the Fermi level was identified in the Ti $L_{2,3}$ SXE spectra of $Ti_3AlC_2$ and $Ti_2AlC$. From these studies, it is clear that the physical and mechanical properties of MAX phases can be further understood from detailed investigations of the underlying electronic structures, and in particular, the M-A and M-X chemical-bond schemes.

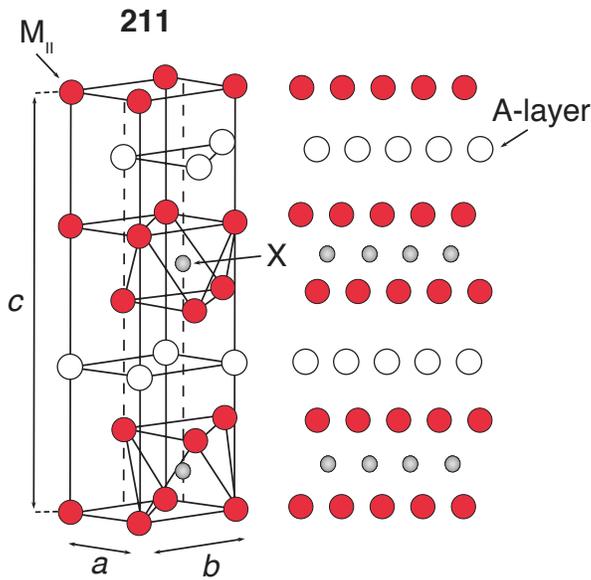

**Figure 1:** The hexagonal crystal structure of 211 ($Ti_2AlN$). There is one A (Al) layer for every second layer of M (Ti) in $Ti_2AlN$. The $M_{II}$ ($Ti_{II}$) atoms have chemical bonds to both X (N) and A (Al) while $M_I$ ($Ti_I$) atoms only bond to N in the case of TiN. The lengths of the measured (calculated) *a* and *c*-axis of the unit cell of $Ti_2AlN$ are 2.98, (3.01) Å and 13.68, (13.70) Å, respectively.

In the present paper, we investigate the electronic structure of the nitrides $Ti_2AlN$ and TiN, using bulk-sensitive and element-specific SXE spectroscopy of single-crystal thin film samples. The SXE technique - with selective excitation energies around the Ti 2*p*, N 1*s*, Al 2*s* and Al 2*p* absorption thresholds - is more bulk sensitive than electron-based spectroscopic techniques. Due to the involvement of both valence and core levels, the corresponding difference in energies of the emission lines and their dipole selection rules, each kind of atomic element can be probed separately. This enables to extract both elemental and chemical bonding information of the electronic structure of the valence bands. The SXE spectra are interpreted in terms of pDOS weighted by the dipole transition matrix elements. The objective of the present investigation is to study the nanolaminated internal electronic structures and the influence of hybridization among the constituent atomic planes in the $Ti_2AlN$ and TiN nitride compounds, in comparison to the isostructural $Ti_2AlC$ and TiC carbide systems with the aim to obtain an increased understanding of the physical and mechanical properties.

# 2  Experimental

## 2.1  Deposition of the Ti$_2$AlN and TiN films

The films were deposited by reactive DC magnetron sputtering from two 3 inch elemental Ti and Al targets in an ultra high vacuum chamber with a base pressure of $\sim 10^{-8}$ Torr. Polished MgO(111) substrates, 10x10x1 mm in size, were used as substrates, cleaned by subsequent ultrasonic baths in trichloroethylene, acetone, and 2-propanol and degassed by holding 900$o$C for one hour prior to deposition. The thickness of the Ti$_2$AlN, respectively TiN was 600 nm, with an initial 120 nm thick TiN(111) seed layer for the Ti$_2$AlN, to prevent Al interdiffusion to the substrate. The depositions were carried out in an Ar/N$_2$ gas mixture of 3.5 mTorr total pressure, with a nitrogen partial pressure of 0.26 mTorr and Ti and Al magnetron powers set to 360 and 100 W, respectively.

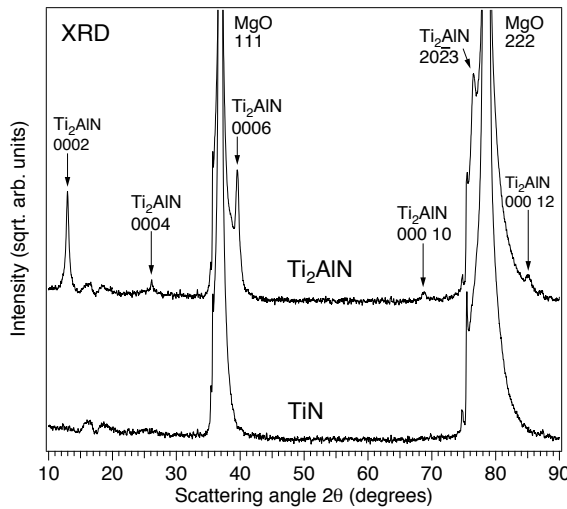

**Figure 2:** X-ray diffractograms from the Ti$_2$AlN(0001) and TiN(111) thin film samples.

The structural properties of the as-deposited films were characterized by x-ray diffraction on a Philips XPert diffractometer using Cu-K$_\alpha$ radiation. The scans in $\Theta$-$2\Theta$-geometry of Ti$_2$AlN (top) and TiN (bottom) are depicted in Fig. 2. Both scans reveal MgO 111 and 222 substrate peaks as denoted. Due to the lattice-matched cube-cube epitaxial growth for TiN, i.e. TiN(111) // MgO(111) and TiN[110] // MgO[110] the TiN peaks for the seed and TiN layer can not be resolved from the MgO substrate peaks. For Ti$_2$AlN we find two competing epitaxial orientations. The main contribution originates from Ti$_2$AlN 000$\ell$ peaks, indicating a parallel basal plane texture with Ti$_2$AlN(0001) // MgO(111) and Ti$_2$AlN[$\bar{1}2\bar{1}0$] // MgO[110]. Another contribution stems from a tilted basal plane orientation, leading to the Ti$_2$AlN 20$\bar{2}$3 peak at a scattering angle of 75.5$o$. The corresponding epitaxial relationship is given by Ti$_2$AlN(20$\bar{2}$3) // MgO(111) and Ti$_2$AlN[$\bar{1}2\bar{1}0$] // MgO[110]. This tilted basal plane growth is induced above a critical thickness, which is subject of ongoing investigations, but does not influence the SXE measurements. The lattice parameters for the films, as determined from reciprocal space maps are a=4.24 Å for TiN and a=2.98 and c=13.68 Å for the Ti$_2$AlN, respectively. The latter are in good agreement with tabulated values of a=2.989 and c=13.614 Å [24]. Chemical analyses by Rutherford backscattering spectroscopy and Elastic recoil detection showed constant elemental distribution over the whole film thickness with compositions according to the formulas given above.

## 2.2 X-ray emission and absorption measurements

The SXE and SXA measurements were performed at the undulator beamline I511-3 at MAX II (MAX-lab National Laboratory, Lund University, Sweden), comprising a 49-pole undulator and a modified SX-700 plane grating monochromator [25]. The SXE spectra were measured with a high-resolution Rowland-mount grazing-incidence grating spectrometer [26] with a two-dimensional multichannel detector with a resistive anode readout. The Ti $L$ and N $K$ SXE spectra were recorded using a spherical grating with 1200 lines/mm of 5 m radius in the first order of diffraction. The Al $L_1$ and $L_{2,3}$ spectra were recorded using a grating with 300 lines/mm, of 3 m radius in the first order of diffraction. The SXA spectra at the Ti 2$p$ and N 1$s$ edges were measured with 0.1 eV resolution using total electron yield (TEY) and total fluorescence yield (TFY), respectively. During the Ti $L$, N $K$, Al $L_1$, $L_{2,3}$ SXE measurements, the resolutions of the beamline monochromator were 0.5, 0.3, 0.2 and 0.01 eV, respectively. The SXE spectra were recorded with spectrometer resolutions of 0.5, 0.3, 0.3 and 0.06 eV, respectively. All measurements were performed with a base pressure lower than $5\times10^{-9}$ Torr. In order to minimize self-absorption effects [27], the angle of incidence was 20$^o$ from the surface plane during the emission measurements. The x-ray photons were detected parallel to the polarization vector of the incoming beam in order to minimize elastic scattering.

# 3 Computational details

## 3.1 Calculation of the x-ray emission spectra

The x-ray emission spectra were calculated within the single-particle transition model by using the augmented plane wave plus local orbitals (APW+lo) band structure method [28]. Exchange and correlation effects were described by means of the generalized gradient approximation (GGA) as parameterized by Perdew, Burke and Ernzerhof [29]. A plane wave cut-off, corresponding to $R_{MT}*K_{max}$=8, was used in the present investigation. For Ti, $s$, $p$ and $d$ local orbitals were added to the APW basis set to improve the convergence of the wave function while, for Al and N only $s$ and $p$ local orbitals were used in their basis set. In order to calculate the Al $L_1$ and Al $L_{2,3}$-edges, the 1$s$, 2$s$ and 2$p$ orbitals of Al were treated as core states, with the 3$s$ and 3$p$ electrons inside the valence shell. The charge density and potentials were expanded up to $\ell$=12 inside the atomic spheres, and the total energy was converged with respect to the Brillouin zone integration.

The x-ray emission spectra were evaluated at the converged ground-state density by multiplying the angular momentum projected density of states by the transition-matrix elements [30]. The electric-dipole approximation was employed so that only the transitions between the core states with orbital angular momentum ℓ to the ℓ±1 components of the electronic bands were considered. The core-hole lifetimes used in the calculations were 0.73 eV, 0.12 eV, 1.3 eV and 0.3 eV for the Ti 2*p*, N 1*s* and Al 2*s*, 2*p* edges, respectively. A direct comparison of the calculated spectra with the measured data was finally achieved by including the instrumental broadening in the form of Gaussian functions corresponding to the experimental resolutions (see experimental section IIB). The final state lifetime broadening was accounted for by a convolution with an energy-dependent Lorentzian function with a broadening increasing linearly with the distance from the Fermi level according to the function $a+b(E-E_F)$, where the constants *a* and *b* were set to 0.01 eV and 0.05 (dimensionless) [31].

### 3.2 Balanced crystal orbital overlap population (BCOOP)

In order to study the chemical bonding of the $Ti_2AlN$ compound, we calculated the BCOOP function by using the full potential linear muffin-tin orbital (FPLMTO) method [32]. In these calculations, the muffin-tin radii were kept as large as possible without overlapping each other (Ti=2.3 atomic units (a.u.), Al=2.35 a.u and N=1.6 a.u.). To ensure a well-converged basis set, a double basis with a total of four different $\kappa^2$ values were used. For Ti, we included the 4*s*, 4*p* and 3*d* as valence states. To reduce the core leakage at the sphere boundary, we also treated the 3*s* and 3*p* core states as semi-core states. For Al, 3*s*, 3*p* and 3*d* were taken as valence states. The resulting basis formed a single, fully hybridizing basis set. This approach has previously proven to give a well-converged basis [33]. For the sampling of the irreducible wedge of the Brillouin zone, we used a special-k-point method [34] and the number of k points were 1000 for $Ti_2AlN$ and 1728 for TiN in the self-consistent total energy calculation. In order to speed up the convergence, a Gaussian broadening of 20 mRy widths was associated with each calculated eigenvalue.

## 4 Results

### 4.1 Ti $L_{2,3}$ x-ray emission

Figure 3 shows Ti $L_{2,3}$ SXE spectra following the $3d4s \rightarrow 2p_{3/2,1/2}$ dipole transitions of $Ti_2AlN$ (full curves) and TiN (dotted curves) excited at 457.0, 462.5 and 490 eV photon energies. For comparison, a Ti $L_{2,3}$ spectrum of pure Ti metal excited at 490 eV is shown by the dashed line. SXA measurements (top, right curves) following the $2p_{3/2,1/2} \rightarrow 3d4s$ dipole transitions were used to locate the energies of the absorption peak maxima at the Ti $2p_{3/2}$ and $2p_{1/2}$ thresholds (vertical ticks). The SXA spectra were normalized to the step edge (below and far above the Ti 2*p* thresholds). The spectra were plotted on a photon energy scale (top) and a relative energy scale (bottom) with respect to the Fermi level ($E_F$). The SXE spectra appear rather delocalized (wide bands) which usually makes electronic structure calculations suitable for the interpretation, particularly for nonresonant spectra. Calculated Ti $L_{2,3}$ spectra of Ti$_2$AlN, TiN and Ti are shown at the bottom of Fig. 3. For comparison of the peak intensities and energy positions, the integrated areas of the experimental and calculated spectra of the three systems were normalized to the calculated Ti $3d+4s$ charge occupations of Ti$_2$AlN: ($3d$: 1.467e, $4s$: 2.026e), TiN ($3d$: 1.429e, $4s$: 2.025e), Ti ($3d$: 1.458e, $4s$: 2.049e). The area for the $L_2$ component was scaled down by the branching ratio and added to the $L_3$ component. For each excitation energy, the spectra were normalized to the time and incoming photon flux by the measured current from a gold mesh in the photon beam.

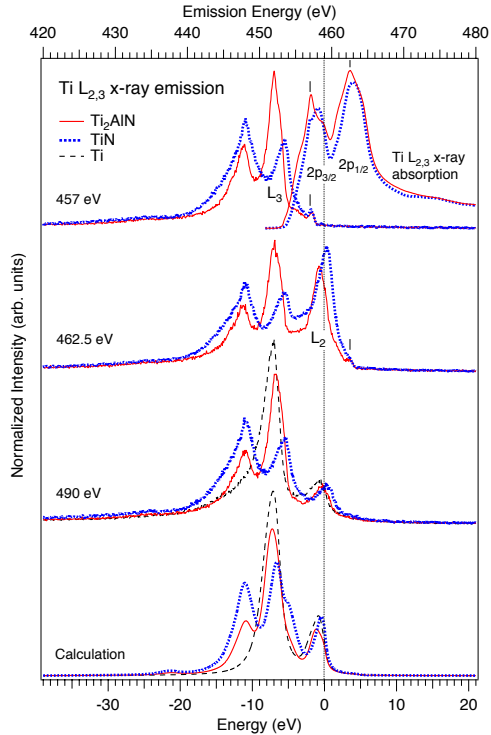

**Figure 3:** Top, Ti $L_{2,3}$ x-ray emission spectra of Ti$_2$AlN and TiN excited at 457.0, 462.5 (resonant) and 490 eV (nonresonant). The pure Ti spectrum was nonresonantly excited at 490 eV. The excitation energies for the resonantly excited emission spectra are indicated by vertical ticks in the x-ray absorption spectra (top, right curves). Bottom, fitted spectra with the experimental $L_{2,3}$ peak splitting of 6.2 eV and the $L_3/L_2$ ratio of 4.2:1.

The calculated spectra consist of the Ti $3d$ and $4s$ pDOS obtained from full-potential *ab initio* density-functional theory projected by the $3d4s \rightarrow 2p$ dipole matrix elements and broadening corresponding to the experimental values. The core-hole lifetime broadening was set to 0.73 eV both for the $2p_{3/2}$ and $2p_{1/2}$ thresholds. To account for the $L_2 \rightarrow L_3M$ Coster-Kronig decay preceding the SXE process [36], increasing the $L_3/L_2$ branching ratio from the statistical ratio (2:1), the calculated spectra were fitted to the experimental nonresonant $L_3/L_2$ ratio of 4.2:1 for Ti$_2$AlN and Ti while it is 2.2:1 for TiN. The observed $L_3/L_2$ ratio (4.2:1) for Ti$_2$AlN and Ti is smaller for the nitrides than for the isostructural carbides (6.0:1) [23] which are both larger than for the more ionic TiN compound (2.2:1). The calculated *ab initio* values of spin-orbit splittings in bandstructure calculations are generally underestimated for the early transition metals (in this case 5.7 eV for Ti $2p$) and overestimated for the late transition metals. The reason for this is not presently known, but must represent effects beyond effective, one-electron theory e.g., many-body effects. In Fig. 3, the fitted $2p_{3/2,1/2}$ spin-orbit splitting was set to the experimental value of 6.2 eV. The energy positions and intensities of the peaks in the fitted spectra of Ti$_2$AlN and TiN are generally in good agreement with the experimental results.

Since our calculations do not include a treatment of polarization effects, we attribute some of the intensity difference to the involvement of the non-spherically symmetric Ti 2*p* core-levels.

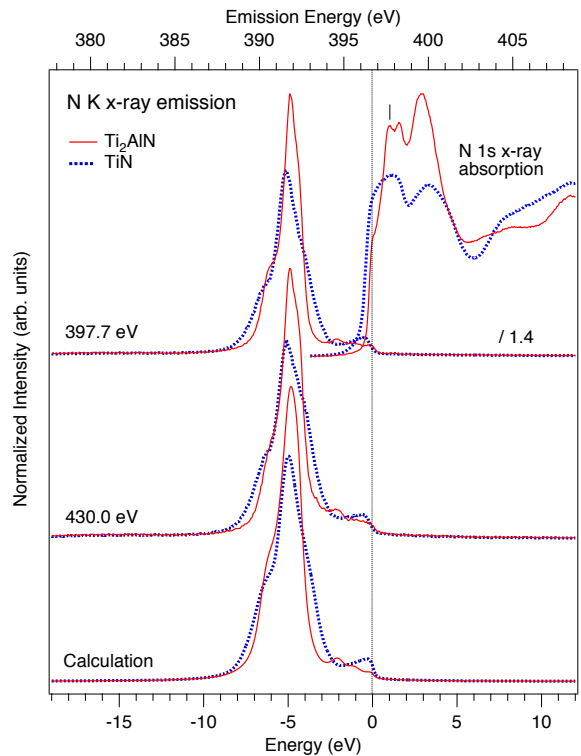

**Figure 4:** Top, experimental N *K* SXE spectra of Ti$_2$AlN and TiN excited at 397.7 eV (resonant) and 430.0 eV (nonresonant), aligned with the N 1*s* core-level XPS binding energy of 396.7 eV for TiN [35]. The resonant excitation energy for the SXE spectra is indicated at the N 1*s* SXA spectra (top, right curves) by the vertical tick. Bottom, calculated emission spectra of Ti$_2$AlN and TiN. The vertical dotted line indicates the Fermi level (E$_F$).

In the spectra excited at 462.5 eV and 490 eV, three peaks are observed at -1 eV, -7 eV and -11 eV, on the relative energy scale at the bottom of Fig. 3. Note that there is a 1.2 eV chemical shift to higher energy for the -1 and -7 eV peak positions in TiN due to the smaller charge occupation and different coordination of the Ti atoms compared to Ti$_2$AlN and Ti. For the -11 eV peak the chemical shift is only 0.3 eV due to its different origin. The -1 eV peak which is not observed in the $L_3$ spectrum excited at 457.0 eV is attributed to Ti $L_2$ emission which is most intense at 462.5 eV excitation energy. The spectral shape of the $L_2$ component is broader and less pronounced than the $L_3$ component due to the larger $2p_{1/2}$ core-hole lifetime broadening.

The -11 eV peak which is absent in the pure metal Ti $L_{2,3}$ spectrum has earlier been interpreted as an intense *anomalous satellite* peak on the low-energy side of the main $L_3$ band in various oxides and nitride compounds [38]. The -11 eV peak is attributed to strong hybridization between the Ti 3*d*4*s* orbitals and the N 2*p* orbitals giving rise to a filled *p−d* band at -4.8 eV below E$_F$. The intensity of the -11 eV peak is ∼ 26% lower in Ti$_2$AlN than in TiN. This is consistent with the observed decrease in stoichiometry when going from TiN to Ti$_2$AlN.

From our bandstructure calculations, we interprete the origin of the -11 eV peak as due to the $L_3$ component of the Ti 3*d* pDOS peak at -4.8 eV below $E_F$ which is shifted -6.2 eV by the 2*p* spin-orbit splitting. The weak $L_2$ component of the 3*d* pDOS contribution (-4.8 eV below $E_F$) overlaps with the much stronger $L_3$ contribution at -7 eV. The weak and broad structure observed in the region -19 to -25 eV with a small peak at -21 eV on the relative energy scale in both Ti$_2$AlN and TiN is due to Ti 3*d* - N 2*s* hybridization at the bottom of the valence band, -12 to -19 eV below $E_F$. Note that this feature is absent in the spectrum of pure Ti.

The origin of the -7 eV peak is related to the $L_3$ component of a series of flat bands of Ti 3*d* character resulting in high pDOS close to the $E_F$, shifted -6.2 eV by the 2*p* spin-orbit splitting. Comparing the Ti$_2$AlN and TiN systems to the corresponding carbide systems, the Ti $L_{2,3}$ peak at -7 eV is absent both in ternary carbide systems when Al has been replaced by Si and Ge [21] and in TiC [23]. On the contrary, the -7 eV peak is strong in both Ti$_2$AlC and Ti$_3$AlC$_2$. This is a signature of relatively strong hybridization between the Ti 3*d* states and the Al states at the top of the valence band. The disappearance of the -7 eV peak in TiC can be explained by the fact that the Ti 3*d* pDOS close to the $E_F$ is very low in TiC, while there is a sharp Ti 3*d* pDOS peak at -2.3 eV below $E_F$. Due to the -6.2 eV 2*p* spin-orbit shift, the main peak is a $L_3$ component appearing at -8.5 eV in the Ti $L_{2,3}$ SXE spectra of TiC. For the ternary carbides, the appearance of the -7 eV peak is thus a signature of Ti 3*d* - Al hybridization, affecting the conductivity and other physical properties, while for the nitrides, the intensity of the -7 eV peak is largely independent of the Al presence.

## 4.2 N *K* x-ray emission

Figure 4 (top) shows N *K* SXE spectra following the 2*p*→1*s* dipole transitions of Ti$_2$AlN and TiN, excited at 397.7 eV (resonant) and 430.0 eV (nonresonant) photon energies. SXA spectra (top, right curves) following the 1*s*→2*p* dipole transitions were measured to identify the absorption maxima and the resonant excitation energy for the SXE spectra. The SXA spectra were normalized to the step edge (below and far above the N 1*s* threshold). Calculated N *K* emission spectra with the N 2*p* pDOS projected by the 2*p*→1*s* dipole transition matrix elements and appropriate broadening corresponding to the experiment are shown at the bottom of Fig. 4. For comparison of the peak intensities and energy positions, the integrated areas of the experimental and calculated spectra of the two systems were normalized to the calculated N 2*p* charge occupations (3.295e for Ti$_2$AlN and 3.303e for TiN). Between the different excitation energies, the spectra were also normalized to the time and incoming photon flux by the measured current from a gold mesh. Thereafter, the intensity of the resonant spectra have been divided by 1.4.

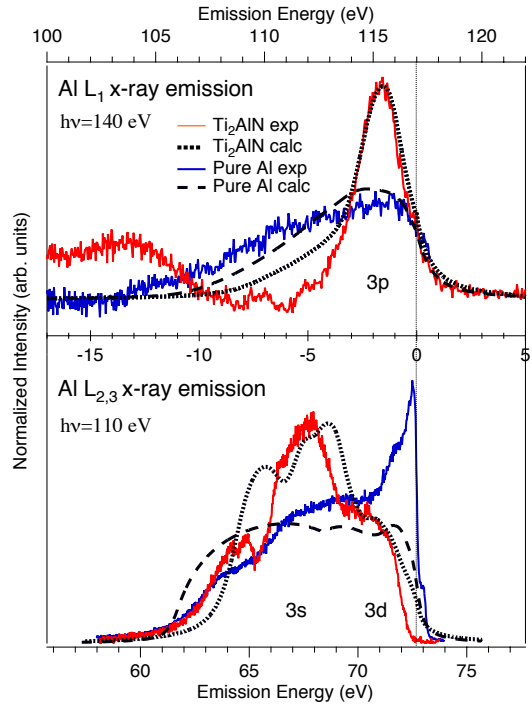

**Figure 5:** Experimental (full curves) and calculated (dotted and dashed curves) Al $L_1$ and Al $L_{2,3}$ SXE spectra of $Ti_2AlN$ and single crystalline Al[100]. The experimental spectra were excited nonresonantly at 140 eV and 110 eV, respectively. The vertical dotted line indicates the Fermi level ($E_F$).

The general agreement between the experimental and theoretical spectra is excellent due to the involvement of the spherically symmetric 1$s$ core levels. The main peak -4.8 eV below $E_F$ has a shoulder on the low-emission energy side at -6 eV below $E_F$ corresponding to a structure in the Ti 3$d$ pDOS. Due to the 25% lower N content, the $Ti_2AlN$ spectra are narrower than for TiN. The $Ti_2AlN$ spectra have an additional peak structure at -2 eV below the $E_F$ attributed to N-Al interaction. Note that the N $K$ intensity and N 2$p$ occupation close to the $E_F$ is lower for $Ti_2AlN$ than for TiN due to the additional interaction with Al, concentrating the bond regions deeper into the valence band. As the excitation energy is changed from resonant (397.7 eV) to nonresonant (430.0 eV), the spectral changes are rather small. For resonant excitation, the -6 eV shoulder is slightly more pronounced in $Ti_2AlN$. The TiN spectra indicate what the N electronic structure of $Ti_2AlN$ would look like if all Al atoms would be exchanged by N atoms. Due to the additional valence electron in N compared to C, the positions of the spectral features related to N are at a lower energy in the nitrides than C in the carbides. This is evident when comparing the N $K$ SXE spectra to the C $K$ SXE of the isostructural carbides, as the C $K$ emission of $Ti_2AlC$ has its main peak at -2.9 eV [23] a shift of +1.9 eV compared to the N $K$ emission of $Ti_2AlN$ (-4.8 eV). The peak shift to lower energy from the $E_F$ in $Ti_2AlN$ indicates stronger interaction and bonding.

### 4.3 Al $L_1$ and $L_{2,3}$ x-ray emission

Figure 5 shows Al $L_1$ (top panel) and Al $L_{2,3}$ (bottom panel) SXE spectra of $Ti_2AlN$ and an Al[100] single crystal, following the $3p \rightarrow 2s$ and $3s,3d \rightarrow 2p_{3/2,1/2}$ dipole transitions, respectively. The measurements were made nonresonantly at 140 eV and 110 eV photon energies. Calculated spectra with the dipole projected pDOS and appropriate broadening are shown by the dotted and dashed curves. A common energy scale with respect to the $E_F$ is indicated in the middle of Fig. 5. For comparison of the peak intensities and energy positions, the integrated areas of the experimental and calculated spectra of the two systems were normalized to the calculated Al $3p$ and $3d+3s$ charge occupations in $Ti_2AlN$: ($3p$: 0.574e, $3d$: 0.063e, $3s$: 0.592e,) and pure Al metal ($3p$: 0.526e, $3d$: 0.090e, $3s$: 0.592e). The area for the $L_2$ component was scaled down by the experimental branching ratio and added to the $L_3$ component.

The general agreement between experiment and theory is better for the $L_1$ emission involving spherically symmetric $2s$ core levels than for the $L_{2,3}$ emission involving $2p$ core levels. Compared to the spectra of pure Al metal, the spectral structures of $Ti_2AlN$ are more focussed to specific energy regions, a few eVs below the $E_F$ as a consequence of bonding to Ti and N. Comparing $Ti_2AlN$ to $Ti_2AlC$ [23], the shift of the N $2p$ orbitals to lower energy in comparison to C $2p$ orbitals (from -2.3 eV to -4.8 eV) implies a shift of the Ti $3d$ pDOS towards lower energy which also affects the spectral distributions of the Al $L_1$ and $L_{2,3}$ spectra. The $L_1$ fluorescence yield is much lower than the $L_{2,3}$ yield making the measurements more demanding. The main Al $L_1$ emission peak in $Ti_2AlN$ at -1.6 eV on the common energy scale is due to Al $3p$ orbitals hybridizing with the Ti $3d$ orbitals. On the contrary, the weak $L_1$ emission of pure Al metal is very broad and flat (0 to -15 eV) without any narrow peak structures, in agreement with our calculated $L_1$ spectrum. However, in the region -3.5 to -10 eV, the intensity of the Al $L_1$ emission is significantly lower in the measured than in the calculated spectrum, indicating that charge is withdrawn and transferred to the N $2p$ and Ti $3d$ orbitals. The small shoulder around -4.8 to -5.0 eV and the valley at -6 eV in the Al $L_1$ emission of $Ti_2AlN$ is mainly caused by hybridization with the N $2p$ orbitals (section IV B). The weak peak structure between -6.8 and -7.7 eV is attributed to hybridization mainly with Ti $3d$ orbitals. The large and broad structure experimentally observed below -9 eV in the $L_1$ spectrum of $Ti_2AlN$ is not reproduced in the calculated $L_1$ spectrum. It can be attributed to hybridization with N $2s$ and Ti $3d$ orbitals at the bottom of the valence band or shake-up transitions in the final state of the emission process [36].

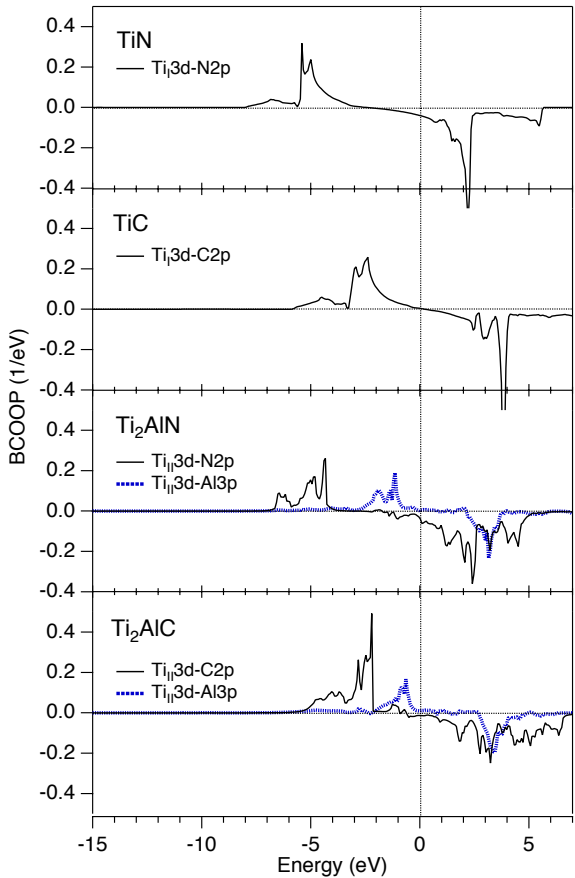

**Figure 6:** Calculated balanced crystal overlap population (BCOOP) of TiN, TiC, $Ti_2AlN$ and $Ti_2AlC$.

The measured Al $L_{2,3}$ SXE spectrum in the lower panel is dominated by $3s \rightarrow 2p_{3/2,1/2}$ dipole transitions while additional $3d \rightarrow 2p_{3/2,1/2}$ transitions mainly occur close to the $E_F$. In particular, this is evident in the Al $L_{2,3}$ spectrum of pure Al metal where a very sharp peak has its maximum at -0.22 eV. The small shoulder at +0.24 eV above $E_F$ is due to Al $L_2$ emission. We find the Al $L_3/L_2$ branching ratio of pure Al metal (4.15:1) to be smaller than in the case of Ti metal (6.3:1). The $2p$ spin-orbit splitting is 0.46 eV, slightly larger than our calculated *ab initio* spin-orbit splitting of 0.44 eV. In contrast to the $L_{2,3}$ SXE spectrum of pure Al metal, the Al $L_{2,3}$ spectrum of $Ti_2AlN$ has a strongly modified spectral weight towards lower emission energy. The main peak has a maximum between -4.8 and -5 eV below $E_F$ and a shoulder at -6 eV, indicating hybridization with the N $2p$ orbitals. The Al $2p$ spin-orbit splitting is not resolved in $Ti_2AlN$. The partly populated $3d$ states are withdrawn from the $E_F$ and form the broad peak structure around -2 eV. For the Al $L_{2,3}$ SXE spectra, the calculated $3s,3d \rightarrow 2p_{3/2,1/2}$ matrix elements are found to play an important role for the spectral shape by reducing the intensity at the bottom of the valence band although this effect is not enough for pure Al metal [37]. The sharp spectral structures at -7.8 and -8.5 eV below $E_F$ in the Al $L_{2,3}$ SXE spectrum of $Ti_2AlN$ can be attributed to hybridized Al $3s$ states with Ti $3d$-orbitals and a valley at -7.4 eV indicates withdrawal of charge in this region.

### 4.4 Chemical Bonding

For $Ti_2AlN$, the equilibrium *a*- and *c*-axis values were calculated to be 3.00 Å and 13.70 Å, respectively. These values are in good agreement with the experimental values of 2.98 and 13.68 Å presented in section II A. In order to analyze the chemical bonding in more detail, we show in Fig. 6 the calculated BCOOP [40] of $Ti_2AlN$ compared to TiN and the corresponding isostructural carbides $Ti_2AlC$ and TiC [23]. The BCOOP makes it possible to compare the strength of two similar chemical bonds where a positive function below $E_F$ means bonding states and a negative function above $E_F$ means anti-bonding states. The strength of the covalent bonding is determined by comparing the integrated areas under the BCOOP curves. Also, an increased energy distance of bonding peak positions from the $E_F$ implies a larger strength of the covalent bonding. The integrated bonding area below $E_F$ in Fig. 6 is ~ 50% larger for TiC than for TiN. However, the distance of the main peak from the $E_F$ is ~ two times larger in TiN (-5.4 eV) in comparison to TiC (-2.6 eV). From this, it can be understood that the covalent $Ti_{II}$ 3*d* - N 2*p* bonding in TiN is stronger than the $Ti_{II}$ 3*d* - C 2*p* bonding in TiC. This is also consistent with the shorter $Ti_{II}$-N bond length in Table I. The 3*d* states in the BCOOP curves in $Ti_2AlN$ are generally located further away from the $E_F$ than in $Ti_2AlC$ which indicates that the $Ti_{II}$-N bond is stronger in $Ti_2AlN$ than the $Ti_{II}$-C bond in $Ti_2AlC$. As the Ti atoms bond stronger to N and C in one direction than to Al in the other direction, the $Ti_{II}$-N and $Ti_{II}$-C bonds are even stronger in $Ti_2AlN$ and $Ti_2AlC$ than the $Ti_I$-N and $Ti_I$-C bonds in TiN and TiC as shown by the shorter bond lengths in Table I.

**Table 1:** Calculated bond lengths [Å] for TiN, TiC, $Ti_2AlN$ and $Ti_2AlC$, where X is either N or C. $Ti_I$ is bonded to X while $Ti_{II}$ is bonded to both X and A as illustrated in Fig. 1.

| Bond type | $Ti_I$ - X | $Ti_{II}$ - X | $Ti_{II}$ - Al | Al - X |
|---|---|---|---|---|
| TiN | 2.129 | - | - | - |
| TiC | 2.164 | - | - | - |
| $Ti_2AlN$ | - | 2.088 | 2.834 | 3.826 |
| $Ti_2AlC$ | - | 2.117 | 2.901 | 3.875 |

The $Ti_{II}$-Al BCOOP peak at -1.1 eV in $Ti_2AlN$ has ~ 15% larger integrated intensity than the corresponding $Ti_{II}$-Al peak at -0.64 eV in $Ti_2AlC$. This shows that the $Ti_{II}$-Al chemical bond in $Ti_2AlN$ is stronger than in $Ti_2AlC$ as also indicated by the shorter bond length in Table I. This is also verified experimentally by the fact that the spectral weight of the Al $L_{2,3}$ SXE spectrum is stronger and slightly shifted away from the $E_F$ in $Ti_2AlN$ in comparison to $Ti_2AlC$ which plays a key role for the physical properties. For the Ti $L_{2,3}$ SXE spectra of $Ti_2AlN$, discussed in section IV A, the BCOOP calculations confirm that the Ti $3d$-N $2p$ hybridization and strong covalent bonding is the origin of the intense Ti pDOS peak at -4.8 eV below the $E_F$ (-11 eV in Fig. 3 when the spin-obit splitting is taken into account). Although a single peak is observed experimentally at -11 eV on the relative energy scale, the BCOOP analysis shows that there are several energy levels in the region between -4 to -7 eV below $E_F$.

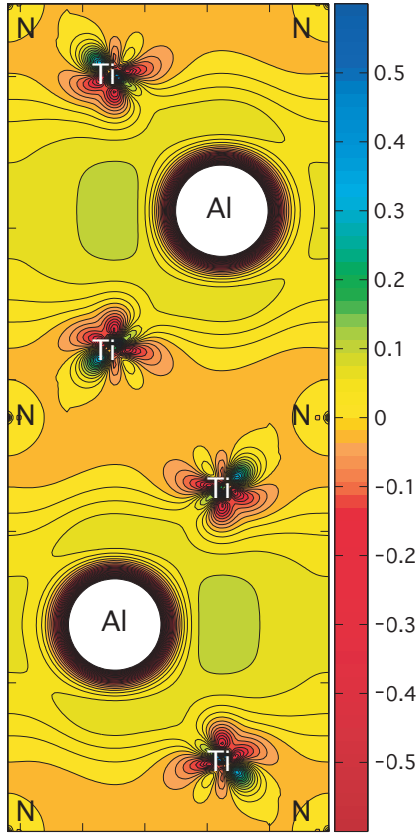

Figure 7 shows a calculated electron density difference plot between $Ti_2AlN$ and $Ti_2N_2$, where in the latter case Al has been replaced by N in the same 211 crystal structure representing a highly twisted TiN cubic structure i.e., $Ti_2N_2$. The plot was obtained by taking the difference between the charge densities of the two systems in the [110] planes of the hexagonal unit cell. Positive values (green/light) mean gain of density and negative values (red/dark) loss of density.

When introducing the Al atoms into the $Ti_2N_2$ matrix we first observe an electron density loss (red/dark colors) at the Al atomic sites since Al atoms have three valence electron, while N have five. Around the Ti atoms, an anisotropic charge density variation is observed with a considerable loss of density (red/dark color). On the other hand, gain of electron density (yellow-green/light color) in the direction towards the N and Al atoms is observed indicating the formation of the Ti-N and Ti-Al bonds. The consequence of the electronic movement is the creation of a certain polarization with a loss of electron density on the neighboring Ti-Ti bonding and therefore reducing its strength. The locally introduced anisotropic electron density distribution around the Ti atoms results in a charge-modulation along the Ti-Al-Ti zigzag bonding direction that propagates throughout the unit cell. The yellow-green/light areas around the N atoms mean a gain of electron density mainly from Ti but also from Al. This shows that the nitrogen atoms respond markedly to the introduction of the Al planes and implies that Al substitution of N results in local modifications to the charge density. Note that in comparison to C in $Ti_2AlC$, N in $Ti_2AlN$ is more electronegative and withdraws a larger part of the electronic density from Al, leading to a stronger Al-N interaction as also indicated by the shorter Al-N bond length in Table I. The charge transfer from Ti and Al towards N is in agreement with the BCOOP presented in Fig. 6.

**Figure 7:** Calculated electron density difference plot between $Ti_2AlN$ and $Ti_2N_2$ (TiN) in the same crystal geometry. Positive values implies gain of density and negative values loss of density [$e/Å^3$]. The plot was obtained by subtracting the charge densities in the [110] diagonal plane of the hexagonal unit cell. The lower valence band energy was fixed to -4.0 Ry (-54.4 eV) and all the Ti $3s^2$, $3p^6$, $3d^2$, $4s^2$, Al $3s^2$, $3p^1$, and N $2s^2$ $2p^3$ valence states were taken into account.

## 5 Discussion

Comparing the crystal structure of $Ti_2AlN$ in Fig. 1 with TiN, it is clear that the physical properties and the underlying electronic structure of the Ti-Al-N system is strongly affected by the intercalated Al layers. The Ti $L_{2,3}$ SXE spectra in Fig. 3 show that the intensity at the $E_F$ is higher in TiN in comparison to $Ti_2AlN$. This is completely opposite to the case for the isostructural carbides $Ti_2AlC$ and TiC [23]. The electrical conductivity/resistivity properties therefore differ significantly between the nitrides and the carbides. Both TiN and $Ti_2AlN$ generally have more dominating Ti $3d$ pDOS at the $E_F$ indicating more metallic-like properties than for the isostructural carbides where the $E_F$ is close to a pronounced pseudogap (a region with low density of states) [23]. The intercalation of Al monolayers into the TiN matrix mainly changes the character of the Ti pDOS close to the $E_F$. Intuitively, the conductivity would increase since Al metal is a good conductor. However, the conductivity is largely governed by the Ti metal bonding and is roughly proportional to the number of states at the Fermi level (TiN: 0.43 states/eV/atom, $Ti_2AlN$: 0.41 states/eV/atom, TiC: 0.12 states/eV/atom and $Ti_2AlC$: 0.34 states/eV/atom).

Experimentally, $Ti_2AlN$ films thus have lower resistivity (0.39 μΩm [8]) compared to $Ti_2AlC$ (0.44 μΩm [41]) while the resistivity of TiN is even lower (0.13 μΩm [42]) and for TiC more than an order of magnitude higher (2.50 μΩm [43]). From our previous 312 study of ternary carbides [23], it was clear that the $Ti_{II}$ layers contribute more to the conductivity than the $Ti_I$ layers. Therefore, one would also expect that $Ti_2AlN$ has higher conductivity than other ternary nitrides since it only contains $Ti_{II}$. Indeed, the resistivity of the other stable nitride system, $Ti_4AlN_3$ is almost an order of magnitude higher [44] than for the $Ti_2AlN$ film. Apart from the covalent contribution, the chemical bonding in binary and ternary carbides and nitrides also has an ionic component. The ionic contribution is expected to be stronger in the nitride systems than in the carbides because of the higher electronegativity of N with respect to C. The latter effect is also observed in the charge density plot (Fig. 7).

From Figs. 3-6, we identified three types of covalent chemical bonds, the strong Ti $3d$ - N $2p$ bond, the weaker Ti $3d$ - Al $3p$ bond and the Ti $3d$ - N $2s$ bond. The Ti $3d$ - N $2p$ and Ti $3d$ - N $2s$ hybridizations are both much deeper in energy from the $E_F$ than the Ti $3d$ - Al $3p$ hybridization indicating stronger bonding. Strengthening the relatively weak Ti $3d$ - Al $3p$ bonding would effectively increase the stiffness of the material. Such a bond strengthening is indeed observed in Ti$_2$AlN in comparison to Ti$_2$AlC causing the $E$-modulus to increase from 260 GPa [10] to 270 GPa [8]. However, the $E$-modulus of both Ti$_2$AlN and Ti$_2$AlC are both significantly lower than for TiN (449 GPa [11]) and TiC$_{0.8}$ (388 GPa [12]).

Although we have shown that the Ti $3d$ - Al $3p$ bonding is slightly stronger in Ti$_2$AlN than in Ti$_2$AlC, the deformation and delamination mechanism is expected to be rather similar in both systems due to the fact that the Ti $3d$ - Al $3p$ bonds are still much weaker in comparison to the Ti$_{II}$$3d$ - C $2p$ and Ti$_{II}$$3d$ - N $2p$ bonds. By choosing C and/or N in the design of the ternary MAX-phases, the physical and mechanical properties can thus be tailored for specific applications. A fractional substitution of C by N in quaternary (pseudo-ternary) MAX-phases allows further fine-tuning of the materials properties, following the evolution of the chemical bonds.

# 6 Conclusions

In summary, we have investigated the electronic structures of Ti$_2$AlN and TiN and compared the results to those of the isostructural Ti$_2$AlC, TiC and pure Ti and Al metals. The combination of soft x-ray emission spectroscopy and electronic structure calculations show that the pronounced peak structures in Ti $L_{2,3}$ x-ray emission have very different spectral intensity weights and energy positions in Ti$_2$AlN and Ti$_2$AlC. This clearly shows the difference in the bond scheme between these two compounds. The Ti $L_3/L_2$ branching ratio is significantly larger in Ti$_2$AlN and Ti than in TiN, indicating metallic properties in the former compounds and more ionic properties in TiN. A strong peak structure in the Ti $L$ emission is observed -4.8 eV below the Fermi level in the Ti $L_{2,3}$ emission and is attributed to intense Ti $3d$ - N $2p$ hybridization and strong covalent bonding while another peak observed -1 eV below the Fermi level is due to Ti $3d$ states hybridized with Al $3p$ states at -1.6 eV in the Al $L_1$ emission in a weaker covalent bonding. In addition, Ti $3d$ - N $2s$ hybridization is identified around -15 eV below the Fermi level as a weak spectral structure in the Ti $L_{2,3}$ emission. Our data of the Al $L_{2,3}$ emission in Ti$_2$AlN as compared to pure Al metal shows a significant shift towards lower energy. This signifies a transfer of charge from the Al $3d$ orbitals towards the Ti and N atoms. The Al $L_{2,3}$ x-ray emission spectrum of Al in Ti$_2$AlN appear very different from the case of Ti$_2$AlC, exhibiting stronger hybridization and interaction between the Al-atoms and Ti and N. The bond regions of Al $3p$ and $3s$ orbitals to Ti $3d$ and N $2p$ orbitals are identified when comparing the Al $L_1$ and $L_{2,3}$ spectra of Ti$_2$AlN to spectra of pure Al metal. The calculated orbital overlaps also show that the Ti $3d$ - Al $3p$ bonding orbitals in Ti$_2$AlN are stronger than in Ti$_2$AlC which implies a change of the elastic properties (higher $E$-modulus) and a higher electrical and thermal conductivity. The analysis of the underlying electronic structure thus provides increased understanding of the chemical trend of materials properties when replacing C by N in Ti$_2$AlC and TiC to Ti$_2$AlN and TiN. Generally, the covalent bonding scheme is important for the understanding of the mechanical and physical properties of these thermodynamically stable nanolaminates. A tuning of the elastic properties and conductivity by alloying or partly exchanging C with N atoms in a material implies that these nanolaminated systems can effectively be tailored during the materials design.

# 7 Acknowledgements

We would like to thank the staff at MAX-lab for experimental support. This work was supported by the Swedish Research Council, the Göran Gustafsson Foundation, the Swedish Strategic Research Foundation (SSF), Strategic Materials Reseach Center on Materials Science for Nanoscale Surface Engineering (MS$^2$E), and the Swedish Agency for Innvovations Systems (VINNOVA) Excellence Center on Functional Nanostructured Materials (FunMat).